\documentclass[leqno,11pt]{article}
\usepackage{amsfonts,latexsym}
\usepackage{amsmath}
\usepackage{amscd}
\usepackage{float,amsmath,amssymb,mathrsfs,bm,multirow,graphics}
\usepackage[dvips]{graphicx}

\newcommand{\nequation}{\setcounter{equation}{0}}

\newcommand{\R}{{\Bbb R}}

\newcommand{\C}{{\Bbb C}}
\newcommand{\Z}{{\Bbb Z}}

\newcommand{\proofbegin}{\noindent{\it Proof.\,\,}}
\newcommand{\proofend}{\hfill$\Box$\bigskip}

\newtheorem{theorem}{Theorem}[section]
\newtheorem{proposition}[theorem]{Proposition}

\newtheorem{figuretext}{Figure}

\input epsf
\begin{document}

\begin{center}

{\LARGE \sc On a novel integrable generalization of the sine-Gordon equation\\} \vspace {7mm}  \noindent

{\large J. Lenells$^{a}$} and {\large A. S. Fokas$^{b}$}

\vskip.7cm

\hskip-.6cm
\begin{tabular}{c}
$\phantom{R^R}^{a}${\small Institut f\"ur Angewandte Mathematik, Leibniz Universit\"at Hannover}\\ 
{\small Welfengarten 1, 30167 Hannover, Germany} \\
{\small E-mail: lenells@ifam.uni-hannover.de} \\
\\
$\phantom{R^R}^{b}${\small Department of Applied Mathematics and Theoretical Physics, University of Cambridge,  }
\\ {\small Cambridge CB3 0WA, United Kingdom} \\
{\small E-mail: t.fokas@damtp.cam.ac.uk} \\
\\
\end{tabular}
\vskip.5cm
\end{center}
\input epsf

\begin{abstract} 

\noindent
We consider an integrable generalization of the sine-Gordon (sG) equation that was earlier derived by one of the authors using bi-Hamiltonian methods. This equation is related to the sG equation in the same way that the Camassa-Holm equation is related to the KdV equation. In this paper we: (a) Derive a Lax pair. (b) Use the Lax pair to solve the initial value problem on the line. (c) Analyze solitons. (d) Show that the generalized sG and sG equations are related by a Liouville transformation. (e) Derive conservation laws. (f) Analyze traveling-wave solutions.
\end{abstract}



%\tableofcontents
\section{Introduction} \nequation
We consider the following integrable generalization of the sine-Gordon equation, which was derived in \cite{F} using bi-Hamiltonian methods:
\begin{equation}\label{gsg}
  u_{tx} = (1 + \nu \partial_x^2)\sin(u), \qquad x \in \R, \quad t > 0,
\end{equation}
where $u(x,t)$ is a scalar-valued function and $\nu \in \R$ is a parameter. 
Equation (\ref{gsg}) is related to the sine-Gordon equation (i.e. to the equation obtained from (\ref{gsg}) by letting $\nu = 0$) in the same way that the Camassa-Hom (CH) equation (see \cite{C-H, F-F}) is related to the Korteweg-de Vries (KdV) equation. Actually, there exist even deeper analogies between (\ref{gsg}) and the CH equation. Indeed, recall that for a particular class of initial conditions it is possible to use a Liouville transformation to map CH to an equation in the KdV hierarchy. Similarly, it will be shown here that for $\nu < 0$, equation (\ref{gsg}) is related to the sine-Gordon equation via an appropriate Liouville type transformation (see Proposition \ref{liouvilleprop}). Furthermore, one of the distinguished features of the CH equation is that it possesses certain non-smooth traveling-wave solutions called peakons. Similarly, it will be shown here that equation (\ref{gsg}), in addition to smooth traveling-wave solutions (see Proposition \ref{smoothprop}), also possesses certain non-smooth traveling-wave solutions called cuspons (see Proposition \ref{cuspprop}). 

Taking into consideration the extensive interest in the literature for the CH equation, as well as the above similarities between (\ref{gsg}) and CH, it is natural to expect that equation (\ref{gsg}) may also be of some interest. 

This paper is organized as follows: Section \ref{laxsec} presents an algorithmic derivation of a Lax pair for equation (\ref{gsg}), starting from the knowledge of the recursion operator (this approach might be useful also for other integrable equations generated via bi-Hamiltonian methods, since for such equations the associated recursion operator is known). Section \ref{specsec} presents the solution of the initial-value problem of equation (\ref{gsg}) with decaying initial conditions in the case that $\nu < 0$. The more interesting case of $\nu > 0$ remains open. Solitons and conservation laws are discussed in sections \ref{solitonsec} and \ref{conssec}. The explicit Liouville transformation relating (\ref{gsg}) with the sine-Gordon equation is presented in section \ref{relationsec}. Smooth, as well as non-smooth traveling-wave solutions are derived in section \ref{travsec}. 

Taking into consideration that (\ref{gsg}) is related to the sG equation with an explicit transformation (for $\nu < 0$), it follows that section \ref{specsec} also presents the solution of the initial-value problem of the sG equation. The solution of this problem was first presented in the classical work \cite{AKNS}, however the method presented in section \ref{specsec} is based on the `modern' Riemann-Hilbert approach and is also influenced by the method of \cite{F1997} (so that it is not necessary to formulate separately the time evolution of the spectral functions, see \cite{FLkdv}).

After this work was completed, the references \cite{BRT, Ra} (where this equation is studied in connection with pseudospherical surfaces) and \cite{SaSa} (where the transformation between (\ref{gsg}) and the sine-Gordon equation is presented) were brought to the authors attention.

\section{A Lax pair} \label{laxsec}\nequation
Let
$$\sigma_1 = \begin{pmatrix} 0	&	1 \\ 1	& 0\end{pmatrix}, \qquad
\sigma_2 = \begin{pmatrix} 0	&	-i \\ i	& 0\end{pmatrix}, \qquad
\sigma_3 = \begin{pmatrix} 1	&	0 \\ 0	& -1\end{pmatrix}.$$
Equation (\ref{gsg}) is the compatibility condition of the following Lax pair:
\begin{equation}\label{lax}
\begin{cases}
	& \psi_x + i \lambda \sigma_3 \psi = W_1 \psi,	\\
	& \psi_t + i\left(\nu\lambda - \frac{1}{4\lambda}\right)\sigma_3 \psi = W_2 \psi,
\end{cases}
\end{equation}	
where $\psi(x, t, \lambda)$ is a $2\times2$-matrix valued function, $\lambda \in \hat{\C} = \C \cup \infty$ is a spectral parameter, and $W_j(x,t,\lambda)$, $j = 1,2$, are defined by
\begin{align} \nonumber
W_1 = & u_x \left(\frac{i}{2}\sigma_1 + \lambda\sqrt{\nu} \sigma_2\right),
	\\ \label{Wdef}
W_2 = & \frac{i}{4 \lambda}\left[\sin(u) \sigma_2 + (\cos(u) - 1)\sigma_3\right]
 + \left[\frac{i\nu}{2}\cos(u) u_x - \frac{\sqrt{\nu}}{2}\sin(u)\right]\sigma_1
	\\ \nonumber
& + \lambda\left[\nu^{3/2}\cos(u) u_x \sigma_2 - i\nu(\cos(u) - 1)\sigma_3\right].
\end{align}

\subsection{Derivation of Lax pair}
Before deriving the Lax pair (\ref{lax}), we first recall the case of the sine-Gordon equation
\begin{equation}\label{sgq}  
  q_{t} = \sin(\partial_x^{-1} q).
\end{equation}
Equation (\ref{sgq}) admits the isospectral problem ($x$-part of a Lax pair)
\begin{equation}\label{sgxpart}
  \psi_x = L\psi, \qquad L =  - i \lambda \sigma_3 + \frac{i}{2} q \sigma_1,
  \end{equation}	
where $\psi(x, \lambda)$ is a vector valued eigenfunction and $\lambda \in \C$ is a spectral parameter, as well as the recursion operator $R$ defined by
$$R = \theta_2\partial_x^{-1}, \qquad \theta_2 = \partial_x^3 + \partial_xq\partial_x^{-1}q\partial_x.$$  
Differentiating equation (\ref{sgxpart}) with respect to $q$ in the direction $w$ (where $\psi$ and $\lambda$ are considered functionals of $q$), we find
\begin{equation}\label{sgxpartw}
\partial_x \psi'(q)[w] = L\psi'(q)[w] - i\int G_\lambda w dx \sigma_3 \psi + \frac{i}{2}w \sigma_1\psi,
\end{equation}
where $G_\lambda$ denotes the gradient of $\lambda$ defined by
$$\lambda'(q)[w] = \int G_\lambda w dx \quad \text{for all $w$}.$$
Let $\psi^*$ be a solution of the adjoint equation
\begin{equation}\label{psistar}
\psi^*_x = -L^A \psi^*,
\end{equation}
where $L^A = L$ denotes the adjoint of $L$ with respect to the inner product
$$(f, g) = \int f^T g dx, \quad \text{i.e.} \quad (f, Lg) = (L^A f, g).$$
Multiplying (\ref{sgxpartw}) by $\psi^{*T}$ from the left and integrating the resulting equation with respect to $x$, we find
\begin{align}\label{sgxpartw2}
\int \psi^{*T} \partial_x \psi'(u)[w] dx = &\int \psi^{*T} L\psi'(u)[w] dx
	\\ \nonumber
& + \int \left[G_\lambda \left(- i\int \psi^{*T}\sigma_3 \psi dx\right) + \frac{i}{2}\psi^{*T}\sigma_1\psi \right]w dx.
\end{align}
The first two terms of (\ref{sgxpartw2}) cancel in view of (\ref{psistar}). Thus, since $w$ is arbitrary,
\begin{equation}\label{sgGlambda}  
  G_\lambda = \frac{\psi^{*T}\sigma_1\psi}{2\int \psi^{*T}\sigma_3 \psi dx}.
\end{equation}
From the bi-Hamiltonian theory, we expect the gradient $G_\lambda$ to satisfy an eigenvalue equation of the form \cite{F-A}
\begin{equation}\label{RAGlambda}  
  R^A G_\lambda = \mu(\lambda) G_\lambda,
\end{equation}
where $\mu = \mu(\lambda)$ is an eigenvalue and $R^A = \partial_x^{-1}\theta_2$ denotes the adjoint of $R$. In fact, letting $\varphi = \psi^{*T}\sigma_1\psi$, we find that $\varphi$ satisfies the equation
$$\left(\frac{\mu(\lambda) \varphi - \varphi_{xx}}{q}\right)_x = q\varphi_x, \qquad \mu(\lambda) = - 4 \lambda^2,$$
which simplifies to (\ref{RAGlambda}) (up to the irrelevant constant in the denominator of (\ref{sgGlambda})). 

Equation (\ref{RAGlambda}) shows how the eigenfunctions of the $x$-part are related to the recursion operator for the sG equation. The eigenfunction $\varphi = \psi^{*T}\sigma_1\psi$ is sometimes referred to as a `squared eigenfunction' since it is bilinear in $\psi^*$ and $\psi$.

We now proceed to the derivation of the Lax pair (\ref{lax}). 
Equation (\ref{gsg}) possesses the recursion operator \cite{F}
$$R = \theta_2\theta_1^{-1},$$
where
$$\theta_1 = \partial_x + \nu \partial_x^3, \qquad \theta_2 = \partial_x^3 + \partial_xu_x\partial_x^{-1}u_x\partial_x.$$  
In order to find a Lax pair of the generalized sine-Gordon equation (\ref{gsg}) we have to trace the steps used for the derivation of (\ref{RAGlambda}) backwards. 
Namely, knowing the recursion operator $R = \theta_2 \theta_1^{-1}$, the goal is to determine the corresponding $x$-part. Thus, we seek an $x$-part such that the gradient $G_\lambda$ of the spectral parameter $\lambda$ of this $x$-part satisfies (\ref{RAGlambda}) with $R^A = \theta_1^{-1}\theta_2$. We let $q = u_x$ and make the ansatz
$$\psi_x = L\psi, \qquad L = - i \lambda \sigma_3 + f(q, \lambda)\sigma_1 + g(q,\lambda) \sigma_2,$$
where $f$ and $g$ are scalar-valued functionals of $q$ and of the $x$-derivatives of $q$ which depend on $\lambda$.
The analog of equation (\ref{sgxpartw}) is
\begin{align}\label{gsgxpartw}
\partial_x \psi'(q)[w] = & L\psi'(q)[w] - i \int G_\lambda w dx \sigma_3 \psi 
+ \frac{\partial f}{\partial \lambda} \int G_\lambda w dx \sigma_1\psi
	\\ \nonumber
& + \frac{\partial g}{\partial \lambda} \int G_\lambda w dx \sigma_2\psi + (D_f w) \sigma_1\psi + (D_g w) \sigma_2\psi,
\end{align}
where the operators $D_f$ and $D_g$ are defined by
$$D_f = \frac{\partial f}{\partial q} + \frac{\partial f}{\partial q_x}\partial_x + \frac{\partial f}{\partial q_{xx}} \partial_x^2 + \cdots, \qquad
D_g = \frac{\partial g}{\partial q} + \frac{\partial g}{\partial q_x}\partial_x + \frac{\partial g}{\partial q_{xx}} \partial_x^2 + \cdots.$$
Letting $\psi^*$ be a solution of (\ref{psistar}), multiplying (\ref{gsgxpartw}) by $\psi^{*T}$ from the left and integrating the resulting equation with respect to $x$, we find the following equation:
\begin{align}\label{gsgxpartw2}\nonumber
\int \psi^{*T} \partial_x \psi'(u)[w] dx =& \int \psi^{*T} L\psi'(u)[w] dx 
	\\
+ \int \biggl[&G_\lambda \left(- i\int \psi^{*T}\sigma_3 \psi dx 
+\int \psi^{*T}  \frac{\partial f}{\partial \lambda} \sigma_1\psi dx
+\int \psi^{*T}  \frac{\partial g}{\partial \lambda} \sigma_2\psi dx\right)
	\\ \nonumber
&+ D_f^A(\psi^{*T}\sigma_1\psi) + D_g^A(\psi^{*T}\sigma_2\psi) \biggr]w dx.
\end{align}
The first two terms of this equation cancel and hence $G_\lambda$ is given by
$$G_\lambda 
= - \frac{D_f^A(\psi^{*T}\sigma_1\psi) + D_g^A(\psi^{*T}\sigma_2\psi)}{- i\int \psi^{*T}\sigma_3 \psi dx 
+\int \psi^{*T}  \frac{\partial f}{\partial \lambda} \sigma_1\psi dx
+\int \psi^{*T}  \frac{\partial g}{\partial \lambda} \sigma_2\psi dx}.$$
Letting
$$f(q, \lambda) = a(\lambda) q, \qquad g(q, \lambda) = b(\lambda) q,$$
where $a(\lambda)$ and $b(\lambda)$ are complex-valued functions of $\lambda$, we find that the function $\varphi$ defined by
$$\varphi := -D_f^A(\psi^{*T}\sigma_1\psi) - D_g^A(\psi^{*T}\sigma_2\psi),$$
satisfies the equation
\begin{equation}\label{muvarphieq}
\left(\frac{\mu(\lambda) (\varphi + \nu\varphi_{xx}) - \varphi_{xx}}{q}\right)_x = q\varphi_x,
\end{equation}
whenever
\begin{equation}\label{muab}
\mu(\lambda) = - \frac{4 \lambda^2}{1 - 4 \nu \lambda^2}, \qquad a = \frac{i}{2}\sqrt{1 + 4b^2 - 4\lambda^2 \nu}.
\end{equation}
Rearranging (\ref{muvarphieq}), we deduce that $G_\lambda$ satisfies (\ref{RAGlambda}) whenever $\mu,a,b$ satisfy (\ref{muab}). Choosing for simplicity $a = i/2$ and $b = \lambda \sqrt{\nu}$, we find
$$\psi_x = L\psi, \qquad L = - i \lambda \sigma_3 + \frac{i}{2}q\sigma_1 + \lambda \sqrt{\nu} q \sigma_2.$$
Having obtained the $x$-part of the Lax pair (\ref{lax}), in order to find the corresponding $t$-part, we make the ansatz
$$\psi_t = M\psi, \qquad M = \sum_{j = -1}^1 \lambda^j (A_j \sigma_1 + B_j \sigma_2 + C_j \sigma_3),$$
where $A_j, B_j, C_j$ are scalar-valued functionals of $q$ independent of $\lambda$. Identifying terms of $O(\lambda^j)$, $j = -1,0,1,2$, in the compatibility equation $L_t - M_x + [L, M] = 0$ and using that $q_t = (1 + \nu\partial_x^2)\sin(\partial_x^{-1}q)$, long but straighforward computations lead to the $t$-part in (\ref{lax}).

\section{Spectral analysis on the line} \label{specsec} \nequation
The Riemann-Hilbert (RH) formalism for integrating a nonlinear evolution equation is based on the construction of eigenfunctions of the associated Lax pair which can be joined together to a bounded and sectionally analytic function on the Riemann sphere of the spectral parameter $\lambda \in \hat{\C}$. 
In order to define eigenfunctions which possess the asymptotics
$$\mu = I + O(1/\lambda), \qquad \lambda \to \infty,$$
we will use a different representation of the Lax pair (\ref{lax}). This representation involves an eigenfunction $\phi$, which is related to $\psi$ via the gauge transformation $\psi = g \phi$. The form of $g$ is such that it (a) diagonalizes the highest order terms in $\lambda$ as $\lambda \to \infty$ and (b) preserves the natural symmetry properties of the Lax pair. 

The spectral analysis of the Lax pair satisfied by $\phi$ is standard: We introduce two particular eigenfunctions via integration from $x = \pm \infty$, and use these eigenfunctions to formulate a $2 \times 2$-matrix RH problem with jump across the real axis. The jump matrix is given in terms of two spectral functions $a(\lambda)$ and $b(\lambda)$, which are defined in terms of the initial data $u_0(x) = u(x,0)$. The possible poles of $a(\lambda)$ give rise to singularities of the RH problem which correspond to solitons. 

The solution $u(x,t)$ can be recovered from the solution of the RH problem, so that this approach provides the solution of the Cauchy problem for equation (\ref{gsg}). However, the RH problem is naturally formulated in terms of variables $(y,t)$ rather than $(x,t)$ where $y$ is a new variable. Therefore, we will only obtain the solution $u(x,t)$ in parametric form. This type of parametric representation arises also in the analysis of the Camassa-Holm and Degasperis-Procesi equations (cf. \cite{CGI, L2002} in the case of the Camassa-Holm equation).

In the rest of the paper we assume that $\nu < 0$ and then for simplicity let $\nu = -1$. 

\subsection{Symmetries}
Assuming that $\nu = -1$, the Lax pair (\ref{lax}) can be written as
\begin{equation}\label{laxLM}
\begin{cases}
	& \psi_x = L \psi,	\\
	& \psi_t = M \psi, 
\end{cases}
\end{equation}	
where
$$L = -i \lambda \sigma_3 + W_1, \qquad
M = i\left(\lambda + \frac{1}{4\lambda}\right)\sigma_3 + W_2,$$
and $W_1$, $W_2$ are given by (\ref{Wdef}) withÊ $\sqrt{\nu}$  replaced by Ê$i$. 
Since $L$ and $M$ are trace-less and satisfy the symmetry relations
$$L^\dagger(\bar{\lambda}) = - L(\lambda), \qquad M^\dagger(\bar{\lambda}) = - M(\lambda),$$
where $A^\dagger$ denotes the complex-conjugate transpose of a matrix $A$, it follows that the functions
$$\det(\psi(x,t, \lambda)) \quad \text{and}\quad \psi^\dagger(x,t,\bar{\lambda})\psi(x,t,\lambda)$$
are independent of $x$ and $t$. We deduce that $\psi$ satisfies 
\begin{equation}\label{psisymm}
\det(\psi(x,t, \lambda)) = 1, \qquad 
\psi^\dagger(x,t,\bar{\lambda}) = \psi^{-1}(x,t,\lambda),
\end{equation}
for all $(x,t)$ provided that these equalities hold at only one point $(x_0, t_0)$. The eigenfunctions introduced in the sequel will indeed have this property. 

\subsection{Eigenfunctions}
Suppose that $u(x,t)$ is a sufficiently smooth real-valued solution of (\ref{gsg}) and suppose that $\cos(u(x,t)) - 1$ has decay as $x \to \pm \infty$ for all $t \geq 0$. Then the form of equation (\ref{gsg}) implies that $\int_\R \sin(u(x,t)) dx = 0$ for each $t$.
Define $m(x,t)$ by
\begin{equation}\label{mdef}
  m(x,t) = 1 + u_x^2(x,t).
\end{equation} 
The gauge transformation
\begin{equation}\label{psiinftydef}
  \psi(x,t,\lambda) = g(x,t) \phi(x,t,\lambda),
\end{equation}  
where
\begin{equation}\label{ginftydef}
  g(x,t) = \sqrt{\frac{1 + \sqrt{m}}{2\sqrt{m}}}\begin{pmatrix} 1 & i\frac{1 - \sqrt{m}}{u_x} \\
i\frac{1 - \sqrt{m}}{u_x} & 1 \end{pmatrix},
\end{equation}
transforms the Lax pair (\ref{lax}) into 
\begin{equation}\label{laxinfty}
\begin{cases}
	& \phi_x + i \lambda p_x \sigma_3 \phi = V_1 \phi,	\\
	& \phi_t  + i (\lambda p_t - \frac{1}{4\lambda})\sigma_3 \phi = V_2 \phi,
\end{cases}
\end{equation}	
where $V_1(x,t)$ and $V_2(x,t,\lambda)$ are defined by
\begin{align*}
V_1 = & \frac{i}{2} \left(u_x + \frac{u_{xx}}{m}\right) \sigma_1,
	\\ 
V_2 = &- \frac{i}{4\lambda}\sigma_3 + i\frac{\cos(u) - u_x\sin(u)}{4\sqrt{m} \lambda} \sigma_3 
+ i\frac{u_x \cos(u) + \sin(u)}{4\sqrt{m} \lambda} \sigma_2 
	\\
& - \frac{i}{2} \left(u_x + \frac{u_{xx}}{m}\right) \cos(u) \sigma_1
\end{align*}
and $p(x,t)$ is a real-valued function defined by
\begin{equation}\label{pdef}
  p(x,t) = x - t + \int_{-\infty}^x (\sqrt{m(x',t)} - 1)dx'. 
\end{equation}
The conservation law
\begin{equation}\label{conslaw}
  \left(\sqrt{m}\right)_t + \left(\cos(u)\sqrt{m}\right)_x = 0
\end{equation}
implies that $p_t = -\cos(u)\sqrt{m}$.
Note that definition (\ref{mdef}) of $m$ implies that $g$ is free of singularities also at points where $u_x = 0$.

The form (\ref{ginftydef}) of $g$ is motivated by the fact that it diagonalizes the terms of $O(\lambda)$ of the Lax pair (\ref{lax}) and that it satisfies
\begin{equation}\label{gsymm}
  \det(g(x,t)) = 1, \qquad g^\dagger(x,t) = g^{-1}(x,t).
\end{equation}  
The relations (\ref{gsymm}) ensure that the gauge transformation (\ref{psiinftydef}) preserves the two properties in (\ref{psisymm}), i.e.
\begin{equation}\label{phisymm}
\det(\phi(x,t, \lambda)) = 1, \qquad 
\phi^\dagger(x,t,\bar{\lambda}) = \phi^{-1}(x,t,\lambda).
\end{equation}
We have defined $V_1$ and $V_2$ so that
$$V_1(x,t) \to 0, \quad V_2(x,t,\lambda) \to 0, \qquad x \to \pm\infty.$$

The form of the Lax pair (\ref{laxinfty}) is convenient for the definition of eigenfunctions which are well-behaved near $\lambda = \infty$. 
Introducing an eigenfunction $\mu$ by
$$\phi = \mu e^{-i(\lambda p - \frac{t}{4\lambda})\sigma_3},$$
we find that the Lax pair (\ref{laxinfty}) can be written as
\begin{equation}\label{mulaxinfty}
\begin{cases}
	& \mu_x + i \lambda p_x [\sigma_3, \mu] = V_1 \mu,	\\
	& \mu_t  + i (\lambda p_t - \frac{1}{4\lambda}) [\sigma_3, \mu] = V_2 \mu.
\end{cases}
\end{equation}	
We define two eigenfunctions $\mu_\pm$ of (\ref{mulaxinfty}) as the solutions of the following two Volterra integral equations
\begin{align} \label{mu+-def}
  \mu_+(x, t, \lambda) = I - \int_x^{\infty} e^{i \lambda (p(x',t) - p(x,t)) \hat{\sigma}_3} V_1(x',t) \mu_+(x',t,\lambda)dx',
 	\\ \nonumber
  \mu_-(x, t, \lambda) = I + \int_{-\infty}^x e^{i \lambda (p(x',t) - p(x,t)) \hat{\sigma}_3} V_1(x',t) \mu_-(x',t,\lambda)dx',
\end{align}
where $\hat{\sigma}_3$ acts on a $2\times 2$ matrix $A$ by $\hat{\sigma}_3A = [\sigma_3, A]$.
The second columns of these equations involve the exponential $e^{2i \lambda (p(x',t) - p(x,t))}$. Since $p(x,t)$ is an increasing function of $x$ for any fixed $t$, we deduce that the second columns of $\mu_+$ and $\mu_-$ are bounded and analytic for $\lambda$ in the upper and lower halves of the complex $\lambda$-plane, respectively. We will use the superscripts $(+)$ and $(-)$ to denote these boundedness properties. Similar considerations are valid for the first columns and hence
$$\mu_+ = \left(\mu_+^{(-)}, \mu_+^{(+)}\right), \qquad
\mu_- = \left(\mu_-^{(+)}, \mu_-^{(-)}\right).$$
The Lax pair (\ref{mulaxinfty}) is of `standard form' as $\lambda \to \infty$, i.e. the highest-order terms of $O(\lambda)$ are diagonal and appear in the exponents of (\ref{mu+-def}), whereas the next order terms of $O(1)$ are off-diagonal. Substitution into (\ref{mulaxinfty}) of the expansion
$$\mu = D + \frac{\mu_1}{\lambda} + \frac{\mu_2}{\lambda^2} + \cdots, \qquad \lambda \to \infty,$$
shows that $D$ is a diagonal matrix independent of $x$ and $t$. In particular, $\mu_+$ and $\mu_-$ have the following asymptotics near $\lambda = \infty$:
\begin{align*}
  \left(\mu_-^{(+)}, \mu_+^{(+)}\right) = I + O(1/\lambda), \qquad \lambda \to \infty, \quad \text{Im}\,\lambda \geq 0,
  	\\
  \left(\mu_+^{(-)}, \mu_-^{(-)}\right) = I + O(1/\lambda), \qquad \lambda \to \infty,
\quad \text{Im}\,\lambda \leq 0.
\end{align*}
Note that $\mu_+$ and $\mu_-$ are nonsingular near $\lambda = 0$, because the integral equations (\ref{mu+-def}) involve only the function $V_1$ and not $V_2$.

\subsection{Spectral functions}
We define the spectral function $s(\lambda)$ by
\begin{equation}\label{lineseq} 
  \mu_+(x,t,\lambda) = \mu_-(x,t,\lambda)e^ {-i(\lambda p(x,t) - \frac{t}{4\lambda})\hat{\sigma}_3} s(\lambda), \qquad \text{Im}\,\lambda = 0.
\end{equation}
Evaluation of (\ref{lineseq}) at $t = 0$ and $x \to -\infty$ gives
\begin{equation}\label{sexplicit}
  s(\lambda) = I - \int_{-\infty}^\infty e^{i \lambda p(x,0) \hat{\sigma}_3} V_1(x, 0) \mu_+(x,0,\lambda)dx, \qquad \text{Im}\, \lambda = 0.
\end{equation}
We deduce from (\ref{phisymm}) that $\mu_\pm$ have the properties
\begin{equation}\label{muinftysymm}
\det(\mu_\pm(x,t, \lambda)) = 1, \quad 
 \mu_{\pm}(x,t,\lambda)_{11} = \overline{\mu_{\pm}(x,t,\bar{\lambda})_{22}}, \quad
 \mu_{\pm}(x,t,\lambda)_{21} = - \overline{\mu_{\pm}(x,t,\bar{\lambda})_{12}}.
\end{equation}
This implies that
$$\det s(\lambda) = 1$$
and that there exist functions $a(\lambda)$ and $b(\lambda)$ such that
$$s(\lambda) = \begin{pmatrix} \overline{a(\bar{\lambda})} & b(\lambda) \\
- \overline{b(\bar{\lambda})} 	&	a(\lambda) \end{pmatrix}.$$
From the explicit expression (\ref{sexplicit}) for $s(\lambda)$ and the fact that the second column of $\mu_+$ is defined and analytic in $\text{Im}\, \lambda > 0$, we find that $a(\lambda)$ has an analytic continuation to the upper half-plane.

\subsection{Residue conditions}
We assume that $a(\lambda)$ has $N$ simple zeros $\{\lambda_j\}_{j = 1}^{N}$ in the upper half-plane. These eigenvalues are either purely imaginary or arise as complex conjugate pairs $(\lambda, -\bar{\lambda})$.
The second column of equation (\ref{lineseq}) is
\begin{equation}\label{linemu2amu1b}
  \mu_+^{(+)} = a\mu_-^{(-)} + b\mu_-^{(+)}e^{-2i(\lambda p - \frac{t}{4\lambda})}, \qquad \text{Im}\, \lambda = 0.
\end{equation}
Applying $\det \left(\mu_-^{(+)}, \cdot\right)$ to this equation and recalling (\ref{muinftysymm}), we find
$$\det \begin{pmatrix} \mu_-^{(+)}(x, t, \lambda) & \mu_+^{(+)}(x, t, \lambda) \end{pmatrix} = a(\lambda), \qquad \text{Im}\, \lambda \geq 0,$$
where we have used that both sides are well-defined and analytic in the upper half-plane to extend the relation to $\text{Im}\, \lambda \geq 0$.
Hence, if $a(\lambda_j) = 0$, then $\mu_-^{(+)}(x, t, \lambda_j)$ and $\mu_+^{(+)} (x, t, \lambda_j)$ are linearly dependent vectors for each $x$ and $t$. It follows that there exist constants $b_j$ such that
\begin{equation}\label{linemubj} 
  \mu_-^{(+)}(x, t, \lambda_j) = b_j e^{2i (\lambda_j p - \frac{t}{4\lambda_j})} \mu_+^{(+)}(x, t, \lambda_j), \qquad x \in \R, \, t > 0.
\end{equation}
Recalling the symmetries in (\ref{muinftysymm}), the complex conjugate of (\ref{linemubj}) is
$$\mu_-^{(-)}(x, t, \bar{\lambda}_j) = -\bar{b}_j e^{-2i (\bar{\lambda}_j p - \frac{t}{4\bar{\lambda}_j})} \mu_+^{(-)}(x, t, \bar{\lambda}_j), \qquad x \in \R, \, t > 0.$$
Consequently, the residues of $\mu_-^{(+)}/a$ and $\mu_-^{(-)}/\bar{a}$ at $\lambda_j$ and $\bar{\lambda}_j$ are
\begin{align*}
& \underset{\lambda_j}{\text{Res}} \frac{\mu_-^{(+)}(x,t, \lambda)}{a(\lambda)}  =  C_j e^{2i (\lambda_j p(x,t) - \frac{t}{4\lambda_j})} \mu_+^{(+)}(x, t, \lambda_j),
	\\
& \underset{\bar{\lambda}_j}{\text{Res}}  \frac{\mu_-^{(-)}(x,t, \lambda)}{\overline{a(\bar{\lambda})}} 
 = - \bar{C}_j e^{-2i (\bar{\lambda}_j p(x,t) - \frac{t}{4\bar{\lambda}_j})} \mu_+^{(-)}(x, t, \bar{\lambda}_j),
\end{align*}
where $\dot{a} = \frac{da}{d\lambda}$ and $C_j = \frac{b_j}{\dot{a}(\lambda_j)}$.

\subsection{Riemann-Hilbert problem}
We now describe how the solution to the initial-value problem for equation (\ref{gsg}) on the line can be expressed in terms of the solution of a $2 \times 2$-matrix Riemann-Hilbert problem. 
Relation (\ref{lineseq}) can be rewritten in the form of the RH problem
$$M_-(x,t,\lambda) = M_+(x, t, \lambda)J(x, t, \lambda), \qquad \text{Im}\, \lambda = 0,$$
where the matrices $M_-$, $M_+$, $J$ are defined by
\begin{align}\label{MplusMminus}
M_+ =& \left(\frac{\mu_-^{(+)}}{a(\lambda)}, \mu_+^{(+)}\right), \qquad \text{Im}\, \lambda \geq 0;
	\\ \nonumber
M_- =& \left(\mu_+^{(-)}, \frac{\mu_-^{(-)}}{\overline{a(\bar{\lambda})}}\right), \qquad \text{Im}\, \lambda \leq 0;
	\\
\label{lineJdef}
J =& \begin{pmatrix} 1	&	-\frac{b(\lambda)}{\overline{a(\bar{\lambda})}}e^{-2i(\lambda p - \frac{t}{4\lambda})} 	\\
-\frac{\overline{b(\bar{\lambda})}}{a(\lambda)} e^{2i (\lambda p - \frac{t}{4\lambda})}  	&	\frac{1}{a(\lambda)\overline{a(\bar{\lambda})}} \end{pmatrix}, \qquad \text{Im}\, \lambda = 0.
\end{align}
The contour for this RH problem is the real axis. 

However, since the jump matrix $J$, in addition to the spectral functions $a(\lambda)$ and $b(\lambda)$, also depends on the function $p(x,t)$, this RH problem cannot be formulated in terms of the initial data alone. We overcome this problem by changing variables
$$(x,t) \mapsto (y, t), \qquad y = p(x,t).$$
Since $J$ only depends on $x$ through $y = p(x,t)$ we can formulate a family of RH problem parametrized by $(y,t)$. From the solution $M(y,t,\lambda)$ of this RH problem, we recover $u(x,t)$ in parametric form. 

Defining
\begin{equation}\label{lineJydef}
J^{(y)}(y, t, \lambda) = \begin{pmatrix} 1	& 
-\frac{b(\lambda)}{\overline{a(\bar{\lambda})}}e^{-2i(\lambda y - \frac{t}{4\lambda})} 	\\
-\frac{\overline{b(\bar{\lambda})}}{a(\lambda)} e^{2i (\lambda y - \frac{t}{4\lambda})}  	&	\frac{1}{a(\lambda)\overline{a(\bar{\lambda})}} \end{pmatrix}, \qquad \text{Im}\, \lambda = 0,
\end{equation}
we have the following result.

\begin{theorem}\label{lineRHtheorem}
Let $u_0(x)$, $x \in \R$, be such that $\cos(u_0(x)) - 1$ has sufficient smoothness and decay as $x\to \pm \infty$ (for details see (\ref{functionspaces}) below) and such that $\int_\R \sin(u_0) dx = 0$. Define $V_{10}(x)$, $m_0(x)$, and $p_0(x)$ by 
$$V_{10} = \frac{i}{2} \left(u_{0x} + \frac{u_{0xx}}{m_0}\right) \sigma_1, \qquad m_0 = 1 + u_{0x}^2, \qquad p_0(x) = x + \int_{-\infty}^x (\sqrt{m_0(x')} - 1)dx'.$$
Let $\mu_+(x,0, \lambda)$ and $\mu_-(x,0, \lambda)$ be the unique solutions of the Volterra linear integral equations (\ref{mu+-def}) evaluated at $t = 0$, with $V_1(x,0) = V_{10}(x)$ and $p(x,0) = p_0(x)$. Define $\{a(\lambda), b(\lambda), C_j\}$ by
\begin{equation}\label{abRH}
  \begin{pmatrix} b(\lambda) \\
 a(\lambda) \end{pmatrix} = [s(\lambda)]_2, \qquad 
   s(\lambda) = I - \int_{-\infty}^\infty e^{i \lambda p_0(x) \hat{\sigma}_3} V_{10}(x) \mu_+(x,0,\lambda)dx,
\end{equation}
and
\begin{equation}\label{CjRH}
  [\mu_-(x, 0, \lambda_j)]_1 = \dot{a}(\lambda_j) C_j e^{2i (\lambda_j p_0(x) - \frac{t}{4\lambda_j})} [\mu_+(x, 0, \lambda_j)]_2,  \qquad j = 1, \dots, N,
 \end{equation}
where $[A]_1$ and $[A]_2$ denote the first and second columns of a $2\times 2$-matrix $A$. 
We assume that $a(\lambda)$ has $N$ simple zeros $\{\lambda_j\}_{j = 1}^{N}$ in the upper half-plane.

Then
\begin{itemize}
\item $a(\lambda)$ is defined for $\text{Im}\, \lambda \geq 0$, analytic in $\text{Im}\, \lambda > 0$, and $a(\lambda) = 1 + O(1/\lambda)$, $\lambda \to \infty$.
\item $b(\lambda)$ is defined for $\lambda \in \R$ and $b(\lambda) = O(1/\lambda)$, $\lambda \to \infty$.
\item $|a(\lambda)|^2 + |b(\lambda)|^2  = 1, \qquad \lambda \in \R$.
\end{itemize}

Suppose there exists a unique solution $u(x,t)$ of equation (\ref{gsg}) with initial data $u_0(x)$ such that $\cos(u(x, t)) - 1$ has sufficient smoothness and decay as $x\to \pm \infty$ for each $t \geq 0$. Then $u(x,t)$ is given in parametric form by
\begin{subequations}\label{linerecoveru}
\begin{align}
  u(X(y,t), t) = 2 \text{\upshape Im}\left(\int_{-\infty}^y \alpha(y',t) dy'\right) + 2n\pi,
\end{align}
where $2n\pi = \lim_{x \to -\infty} u_0(x)$, the function $X(y,t)$ is defined by
\begin{align}
 X(y,t) = y + t +  \int_{-\infty}^y \left(\sqrt{1 - 4\text{\upshape Im}(\alpha(y',t))^2} - 1\right)dy',
\end{align}
the function $\alpha(y,t)$ is the unique solution of the Ricatti equation
\begin{align}\label{ricatti}
 \alpha_y =  \alpha^2 - 4 i  \left(\lim_{\lambda \to \infty} \lambda M_{12}(y,t,\lambda)\right)\alpha - \frac{1}{4}, \qquad \lim_{y \to \pm \infty} \alpha(y,t) = -\frac{1}{2},
\end{align}
\end{subequations}
which satisfies
$$|\alpha(y,t)| = \frac{1}{2}, \qquad y \in \R, \quad t \geq 0,$$
and $M(y, t, \lambda)$ is the unique solution of the following RH problem:
\begin{itemize}
\item $M(y,t,\lambda) = \left\{ \begin{array}{ll}
M_+(y,t,\lambda), \qquad \text{\upshape Im}\, \lambda \geq 0, \\
M_-(y,t,\lambda), \qquad \text{\upshape Im}\, \lambda \leq 0, \\
\end{array} \right.$ 
is sectionally meromorphic.

\item $M_-(y,t,\lambda) = M_+(y, t, \lambda)J^{(y)}(y, t, \lambda)$ for $\lambda \in \R,$ where $J^{(y)}$ is defined in (\ref{lineJydef}).

\item $M(y, t, \lambda)$ has the asymptotic behavior
\begin{equation}\label{MtoI}
M(y, t, \lambda) = I + O\left(\frac{1}{\lambda}\right), \qquad \lambda \to \infty.
\end{equation}

\item The first column of $M_+$ has simple poles at $\lambda = \lambda_j$, $j = 1, \dots, N$, and the second column of $M_-$ has simple poles at $\lambda = \bar{\lambda}_j$, $j = 1, \dots, N$.
The associated residues are given by
\begin{subequations}\label{lineresidue}
\begin{align}\label{lineresidue1}
& \underset{\lambda_j}{\text{\upshape Res}} [M(y,t,\lambda)]_1  =  C_j e^{2i (\lambda_j y - \frac{t}{4\lambda_j})} [M(y,t,\lambda_j)]_2,
		\\\label{lineresidue2}
& \underset{\bar{\lambda}_j}{\text{\upshape Res}}  [M(y,t,\lambda)]_2
 = - \bar{C}_j e^{-2i (\bar{\lambda}_j y - \frac{t}{4\bar{\lambda}_j})} [M(y,t, \bar{\lambda}_j)]_1.
\end{align}
\end{subequations}
\end{itemize}
\end{theorem}
\proofbegin
It remains to prove (\ref{linerecoveru}). By substituting the expansion
$$\mu(x,t,\lambda) = I + \frac{\mu^{(1)}(x,t)}{\lambda} + \frac{\mu^{(2)}(x,t)}{\lambda^2} + O\left(\frac{1}{\lambda^3}\right), \qquad \lambda \to \infty,$$
into the $x$-part of (\ref{mulaxinfty}), we find by considering the terms of $O(1)$ that
\begin{equation}\label{m112xt}  
   4 \mu^{(1)}_{12}(x,t) = \frac{1}{\sqrt{m(x,t)}}\left(u_x(x,t) + \frac{u_{xx}(x,t)}{m(x,t)}\right).
\end{equation}
Define $Q(y,t)$ by $Q_y(p(x,t),t) = -4 \mu^{(1)}_{12}(x,t)$. Then, by construction of the RH problem,
\begin{equation}\label{QlimitM12}
  Q_y(y,t) = -4 \lim_{\lambda \to \infty} \left( \lambda M_{12}(y,t,\lambda)\right).
\end{equation}
Equation (\ref{m112xt}) is conveniently expressed in terms of $y = p(x,t)$ and $Q$. Indeed, using that $\partial_x = \sqrt{m}\partial_y$ and $m = 1/(1 - u_y^2)$, we find
\begin{equation}\label{Qyuy}  
  Q_y =  -\left(u_y + \frac{u_{yy}}{\sqrt{1 - u_y^2}}\right).
\end{equation}
Integration of this equation yields
\begin{equation}\label{Qurelation}
  Q = -u - \arcsin(u_y),
\end{equation}
where we have fixed the integration constant by assuming that
$$Q(y,t) + u(y,t) \to 0, \qquad y \to \pm \infty.$$
Note that 
$$-1 \leq u_y = \frac{u_x}{\sqrt{1 + u_x^2}} \leq 1,$$
so that $Q$ is well-defined by (\ref{Qurelation}).
We rewrite (\ref{Qurelation}) as
\begin{equation}\label{uysincos}
  u_y = -\sin(Q + u).
\end{equation}  
Defining $\alpha(y,t)$ by 
\begin{equation}\label{alphadef}  
  \alpha = -\frac{1}{2} e^{i(Q + u)},
\end{equation}
equation (\ref{uysincos}) becomes the Ricatti equation 
\begin{equation}\label{ricattiQy}
  \alpha_y =  \alpha^2 + iQ_y \alpha - \frac{1}{4}, \qquad \alpha(\pm \infty,t) = -\frac{1}{2}.
\end{equation}  
By (\ref{uysincos}) and (\ref{alphadef}), 
\begin{equation}\label{eiuuy}
  u_y = 2\text{Im}(\alpha).
\end{equation}

On the other hand, $y \to x - t$ as $x \to -\infty$, and so
\begin{equation}\label{xyrelation}
  x = y + t + \int_{-\infty}^y \left(\sqrt{1 - u_y^2} - 1\right)dy'.
\end{equation}
The parametric formula (\ref{linerecoveru}) now follows from equations (\ref{QlimitM12}), (\ref{ricattiQy}), (\ref{eiuuy}), and (\ref{xyrelation}).
\proofend

The jump matrix $J^{(y)}$ defined in (\ref{lineJydef}) satisfies the symmetry relation $\overline{J_{12}^{(y)}(\bar{\lambda})} = J_{21}^{(y)}(\lambda)$. This implies that there exists a `vanishing lemma' for the RH problem. This guarantees existence in $H_1(\R)$ \cite{FI}. Thus the initial conditions must be chosen in an appropriate function space such that 
\begin{equation}\label{functionspaces}
  p_0(x) - x \in L_\infty, \qquad V_{10} \in L_1, \qquad a, b \in H_1.
\end{equation}

\section{Soliton solutions}\label{solitonsec} \nequation
The solitons correspond to spectral data $\{a(\lambda), b(\lambda), C_j\}$ for which $b(\lambda)$ vanishes identically. In this case the jump matrix $J^{(y)}$ in (\ref{lineJydef}) is the identity matrix and the RH problem of Theorem \ref{lineRHtheorem} consists of finding a meromorphic function $M(y,t,\lambda)$ satisfying (\ref{MtoI}) and the residue conditions (\ref{lineresidue}). In what follows we will suppress the $(y,t)$-dependence of $M$ and write $M(\lambda)$ for $M(y,t,\lambda)$.

Let $\theta_j = \lambda_j y - \frac{t}{4\lambda_j}$. From (\ref{MtoI}) and (\ref{lineresidue1}) we find
\begin{equation}\label{M1decompose}
[M(\lambda)]_1 = \begin{pmatrix} 1	\\	0 \end{pmatrix} + \sum_{j = 1}^{N} \frac{C_j e^{2i\theta_j} [M(\lambda_j)]_2}{\lambda - \lambda_j}.
\end{equation}
It follows from (\ref{muinftysymm}) that $M(\lambda)$ respects the symmetries
\begin{align}\label{Msymmetry}
& M_{11}(\lambda) = \overline{M_{22}(\bar{\lambda})}, \qquad M_{21}(\lambda) = - \overline{M_{12}(\bar{\lambda})}.
\end{align}
Moreover, if (\ref{Msymmetry}) holds, then only one of the two residue conditions (\ref{lineresidue1})-(\ref{lineresidue2}) needs to be verified since the other condition is a consequence of symmetry.
In view of (\ref{Msymmetry}), equation (\ref{M1decompose}) can be written as
\begin{equation}\label{M2kfromM2kj}
\begin{pmatrix} \overline{M_{22}(\bar{\lambda})} \\ -\overline{M_{12}( \bar{\lambda})} \end{pmatrix} = \begin{pmatrix} 1	\\	0 \end{pmatrix} + \sum_{j = 1}^{N} \frac{C_j e^{2i\theta_j}}{\lambda - \lambda_j}\begin{pmatrix} M_{12}(\lambda_j) \\ M_{22}(\lambda_j) \end{pmatrix}.
\end{equation}
Evaluation at $\bar{\lambda}_n$ yields 
\begin{equation}\label{algebraicsystem}
\begin{pmatrix} \overline{M_{22}(\lambda_n)} \\ -\overline{M_{12}(\lambda_n)} \end{pmatrix} = \begin{pmatrix} 1	\\	0 \end{pmatrix} + \sum_{j = 1}^{N} \frac{C_j e^{2i\theta_j}}{\bar{\lambda}_n - \lambda_j}\begin{pmatrix} M_{12}(\lambda_j) \\ M_{22}(\lambda_j) \end{pmatrix}, \qquad n =1, \dots, N.
\end{equation}
Solving this algebraic system for $M_{12}(\lambda_j)$ and $M_{22}(\lambda_j)$, $j = 1, \dots, N$, and substituting the resulting expression into (\ref{M1decompose}) and (\ref{M2kfromM2kj}) gives explicit expressions for the entries of $M(\lambda)$. The $N$-soliton solution $u(x,t)$ is then obtained from (\ref{linerecoveru}). 

\subsection{One-solitons}
The one-solitons arise when $a(\lambda)$ has one purely imaginary zero $\lambda_1$. Letting $N = 1$ in (\ref{algebraicsystem}) and taking the complex conjugate of the second row, we find
\begin{align*}
 \overline{M_{22}(\lambda_1)} = 1 + \frac{C_1 e^{2i\theta_1}}{\bar{\lambda}_1 - \lambda_1} M_{12}(\lambda_1),
\qquad  -M_{12}(\lambda_1) =  \frac{\bar{C}_1 e^{-2i\bar{\theta}_1}}{\lambda_1 - \bar{\lambda}_1} \overline{M_{22}(\lambda_1)}.
\end{align*}
Solving these equations for $M_{12}(\lambda_1)$ and $\overline{M_{22}(\lambda_1)}$, and substituting the result into (\ref{M2kfromM2kj}), we find
$$M_{12}(\lambda) =  -\frac{4 \bar{C}_1 \text{Im}(\lambda_1)^2}{(\lambda - \bar{\lambda}_1) (|C_1|^2 e^{2 i \theta_1} + 4 e^{2 i \bar{\theta}_1} \text{Im}(\lambda_1)^2)}.$$
Equation (\ref{QlimitM12}) yields
$$Q_y = \frac{16 \bar{C}_1 \text{Im}(\lambda_1)^2}{|C_1|^2 e^{2 i \theta_1} + 4 e^{2 i \bar{\theta}_1} \text{Im}(\lambda_1)^2},$$
where $Q(y,t)$ is defined by (\ref{Qurelation}). Letting $\lambda_1 = i \alpha$ and $C_1 = 2 \alpha e^{x_0}$, where $\alpha > 0$ and $x_0 \in \R$ are parameters, we find a real-valued solution given by
\begin{equation}\label{Qonesol}
  Q(y,t) = 4\arctan\left(e^{-2 \alpha y - \frac{t}{2\alpha} + x_0}\right).
\end{equation}
The solution $u(x,t)$ is given implicitly in terms of $Q(y,t)$ by (\ref{linerecoveru}).
We recognize $Q(y,t)$ as a one-soliton solution of the sine-Gordon equation $Q_{ty} = \sin(Q)$. We will see in the next section that this is no coincidence.

\section{Relation to sine-Gordon}\label{relationsec} \nequation
The sine-Gordon equation 
  \begin{equation}\label{sg}
  Q_{\tau y} = \sin(Q)
\end{equation}  
admits the Lax pair
\begin{equation}\label{sglax}
\begin{cases}
	& \phi_y + i \lambda \sigma_3 \phi = -\frac{i}{2}Q_y \phi,	\\
	& \phi_\tau  - \frac{i}{4\lambda}\sigma_3 \phi = \frac{i}{4\lambda} \bigl((\cos(Q) - 1)\sigma_3 -\sin(Q) \sigma_2\bigr)\phi,
\end{cases}
\end{equation}	
where $\phi(y, \tau,\lambda)$ is a $2 \times 2$-matrix valued eigenfunction. On the other hand, the change of variables $(x,t) \mapsto (y, \tau)$ with $y = p(x,t)$ and $\tau = t$ transforms the $x$-part of (\ref{laxinfty}) into
$$ \phi_y + i \lambda \sigma_3 \phi = \frac{i}{2 \sqrt{m}} \left(u_x + \frac{u_{xx}}{m}\right) \sigma_1 \phi.$$
This $x$-part coincides with that of (\ref{sglax}) if we identify $Q_y$ with
$$-\frac{1}{ \sqrt{m}} \left(u_x + \frac{u_{xx}}{m}\right) = -\left(u_y + \frac{u_{yy}}{\sqrt{1 - u_y^2}}\right).$$
This is the identification made in (\ref{Qyuy}) and suggests that if $u(x,t)$ satisfies (\ref{gsg}), then the corresponding function $Q(y, \tau)$ evolves according to an equation in the sine-Gordon hierarchy. In fact, the following proposition shows that $Q(y, \tau)$ evolves exactly according to the sG equation.

\begin{proposition}\label{liouvilleprop}
  Suppose $u(x,t)$ satisfies the generalized sG equation (\ref{gsg}). Then the function $Q(y,\tau)$ defined by
  \begin{subequations}
  \begin{align}\label{Qytau}
  & Q(y,\tau) = -u(x,t) - \arcsin\left(\frac{u_x(x,t)}{\sqrt{1 + u_x^2(x,t)}} \right),
  	\\ \label{ytaudef}
& y = x - t + \int_{-\infty}^x (\sqrt{1 + u_x^2(x',t)} - 1)dx', \qquad \tau = t,	 
  \end{align}
  \end{subequations}
  satisfies the sG equation (\ref{sg}).
\end{proposition}
\proofbegin
In view of the conservation law (\ref{conslaw}), we have the relations
\begin{equation}\label{partialxyt}
\frac{\partial}{\partial t} =  \frac{\partial}{\partial \tau} - \cos(u)\sqrt{m} \frac{\partial}{\partial y},\qquad
\frac{\partial}{\partial x} = \sqrt{m} \frac{\partial}{\partial y}.
\end{equation}
Using these relations we find that the generalized sG equation (\ref{gsg}) in terms of the variables $(y,\tau)$ takes the form
\begin{equation}\label{gsgyt}
   u_{y\tau} = \sqrt{1 - u_y^2} \sin(u).
\end{equation}

Assume that $u(x,t)$ satisfies (\ref{gsg}). Applying $\partial_\tau \partial_y$ to (\ref{Qytau}) and using (\ref{gsgyt}) and its $y$-derivative to replace $u_{yy\tau}$ and $u_{y\tau}$ in the resulting equation, we find
$$Q_{y\tau} = -u_y \cos(u) - \sqrt{1 - u_y^2}\sin(u).$$
Replacing $u$ with $-Q  - \arcsin(u_y)$ in this equation and simplifying, we find that the right-hand side equals $\sin(Q)$.
\proofend

\section{Conservation laws}\label{conssec} \nequation
An infinite sequence of conservation laws can be derived for a bi-Hamiltonian equation by means of the Lenard scheme. This approach however involves the inversion of the bi-Hamiltonian operators. In the case of equation (\ref{gsg}) the operators $\theta_1$ and $\theta_2$ are nonlocal and difficult to invert. Instead of following this procedure we can derive conservation laws of (\ref{gsg}) by utilizing the connection with the sG equation established in the previous section. 
Indeed, suppose that the sG equation (\ref{sg}) admits a conservation law of the form
\begin{equation}\label{AtauBy}  
  A_\tau = B_y,
\end{equation}
where $A$ and $B$ are functionals of $Q(y,\tau)$ and its $y$-derivatives. The change of variables (\ref{ytaudef}) transforms (\ref{AtauBy}) into
\begin{equation}\label{AtBx}  
  \left(\sqrt{m}A\right)_t = \left(B - \cos(u)\sqrt{m} A\right)_x.
\end{equation}
Writing this equation in terms of $u(x,t)$ by means of relation (\ref{Qytau}), we find a conservation law for equation (\ref{gsg}). Since the conservation laws of sG can be constructed explicitly, this approach yields an infinite number of conservatin laws for (\ref{gsg}).
We next illustrate this method by deriving a few conservation laws explicitly.

As noted in section \ref{laxsec}, equation (\ref{sg}) admits the recursion operator $J_2J_1^{-1}$ where the Hamiltonian operators are
$$J_1 = \partial_y, \qquad J_2 = \partial_y^3 + \partial_y q\partial_y^{-1}q\partial_y,$$
and $q = Q_y$. The sG equation is given by
$$q_\tau = J_1 \frac{\delta H_0}{\delta q},$$
where the Hamiltonian $H_0$ is defined by
$$H_0 = \int \left(\cos(\partial_y^{-1}q) - 1\right) dy.$$
Further conservation laws can be determined according to the equation
$$ \frac{\delta H_{n + 1}}{\delta q} = J_1^{-1} J_2 \frac{\delta H_n}{\delta q}.$$
Since $J_1^{-1}$ is given by simple integration, the conserved quantities $H_n$ for $n \geq 0$ can be constructed recursively.
We let $H_n = \int A_n dy$ and write the associated conservation law in the form
$$H_n : (A_n)_\tau = (B_n)_y.$$
We find the following explicit expressions for the first few conservation laws expressed in terms of $Q$:
\begin{align*}
&H_0:  \left(\cos{Q} - 1 \right)_\tau = \left(- \frac{1}{2} (\partial_y^{-1}\sin{Q})^2 \right)_y,
	\\
&H_1: \left( \frac{1}{2} Q_y^2 \right)_\tau = \left( -\cos{Q} \right)_y,
	\\
&H_2: \left(\frac{1}{4}Q_y^4 - Q_{yy}^2 \right)_\tau = \left(-Q_y^2 \cos{Q} \right)_y,
	\\
&H_3: \left( \frac{1}{8}Q_y^6 - \frac{5}{2}Q_y^2Q_{yy}^2 + Q_{yyy}^2 \right)_\tau = \left( -\frac{3}{4} Q_y^4\cos(Q) - 2Q_y^2Q_{yy}\sin(Q) + Q_{yy}^2\cos(Q) \right)_y.
\end{align*}
We use (\ref{Qytau}), (\ref{partialxyt}), and (\ref{AtBx}) to rewrite these conservation laws in terms of $u(x,t)$.
The result for $H_0$ and $H_1$ is
\begin{align*}
H_0: & \left(\cos(u) + (\cos{u})_x - \sqrt{m} \right)_t 
	\\
&= \left( -\frac{1}{2} \left(\partial_x^{-1}(\sin{u} + (\sin{u})_x\right)^2 + \left(\cos(u) + (\cos{u})_x - \sqrt{m}\right)\cos{u} \right)_x,
	\\
H_1:& \left( \frac{(m u_x+u_{xx})^2}{2 m^{5/2}} \right)_t 
	\\
& = \left( \frac{\left(-m \left(m^2+m+m_x\right)-u_{xx}^2\right) \cos
   (u(x,t))+2 m^2 u_x \sin (u(x,t))}{2 m^{5/2}} \right)_x.
\end{align*}
For $H_2$ and $H_3$ the expressions are quite long, thus we only present the conserved quantities:
\begin{align*}
H_2 = & \int \frac{1 }{8 m^{3/2}}\left[\left(\frac{u_{xx}}{m}+u_x\right)^4-\frac{(3 m_x
   u_{xx}-2 m (u_{xx}+u_{xxx}))^2}{m^4}\right] dx,
	\\
H_3 =& \int \frac{1}{16 m^{5/2}}\biggl[-\frac{5 (m u_x+u_{xx})^2 (3 m_x u_{xx}-2 m
   (u_{xx}+u_{xxx}))^2}{m^6} +\left(\frac{u_{xx}}{m}+u_x\right
   )^6
   \\
   &	+\frac{8 \left(m^2
   (u_{xxx}+ u_{xxxx})+m \left(-2 m_x u_{xx}-5 m_x
   u_{xxx}+15 u_{xx}^3\right)-18
   u_{xx}^3\right)^2}{m^6}\biggr] dx.
\end{align*}
It can be verified directly that these indeed are conservation laws for equation (\ref{gsg}).

\section{Traveling-wave solutions}\label{travsec} \nequation
For a traveling wave $u(x,t) = \varphi(x - ct)$ traveling with speed $c$, equation (\ref{gsg}) takes the form
\begin{equation}\label{trav1}
  (\cos(\varphi)-c)\varphi_{xx} = \sin(\varphi) + \varphi_x^2\sin(\varphi).
\end{equation}  
We rewrite (\ref{trav1}) as
$$\frac{\varphi_{xx}}{1 + \varphi_x^2} = \frac{\sin(\varphi)}{\cos(\varphi) - c}.$$
Multiplying this equation by $\varphi_x$ and integrating with respect to $x$, we find
\begin{equation}\label{logvarphi}
  \frac{1}{2}\log(1 + \varphi_x^2) = - \log|c -\cos(\varphi)|+ \log(b),
\end{equation}
where $b > 0$ is an integration constant. Exponentiating and then taking the square of both sides of equation (\ref{logvarphi}), we find after rearrangement
\begin{equation}\label{travODE}
   \varphi_x^2 = F(\varphi), \qquad F(\varphi) = \frac{b^2 - (c - \cos(\varphi))^2}{(c - \cos(\varphi))^2}.
\end{equation}
The qualitative structure of the traveling-wave solutions of (\ref{gsg}) can be obtained by analyzing this ordinary differential equation. It is in fact possible to solve (\ref{travODE}) explicitly, which yields an expression for $\varphi^{-1}$ in terms of elliptic functions. We proceed instead by briefly describing the qualitative structure of a few different types of traveling waves. We refer to \cite{Ltrav} for a more detailed analysis in the similar case of the Camassa-Holm equation.

The qualitative structure of solutions of (\ref{travODE}) is determined by the zeros and the poles of $F(\varphi)$. 
The sets of zeros and poles of $F(\varphi)$ are given by
\begin{equation*}\label{zeropoleset}
\{\varphi | \cos(\varphi) = c \pm b\}\quad \text{and}\quad  \{\varphi | \cos(\varphi) = c\},
\end{equation*}
respectively. Since $b > 0$, these sets are disjoint.

The traveling waves are parametrized by $b > 0$ and $c\in \R$. Since we are interested in waves such that $\lim_{x \to \pm \infty} \varphi(x) \in 2\pi \Z$ we assume that $b = |c - 1|$. In this case $F(\varphi)$ has double zeros at $\varphi = 2n\pi$, $n \in \Z$, and these zeros give rise to solutions $\varphi(x)$ of (\ref{travODE}) which approach elements in the set $2\pi \Z$ exponentially fast as $x \to \pm \infty$. In addition to these solutions there exists a range of periodic solutions obtained for values of $b$ such that $b \neq |c - 1|$. 

Note that if $u(x,t)$ solves equation (\ref{gsg}), then so do the functions $(x,t)\mapsto \pm u(x,t) + 2\pi n$, $n \in \Z$. This corresponds to the fact that $F(\varphi)$ is an even function of period $2\pi$.

\subsection{Smooth waves}
\begin{figure}
\begin{center}
    \includegraphics[width=.5\textwidth]{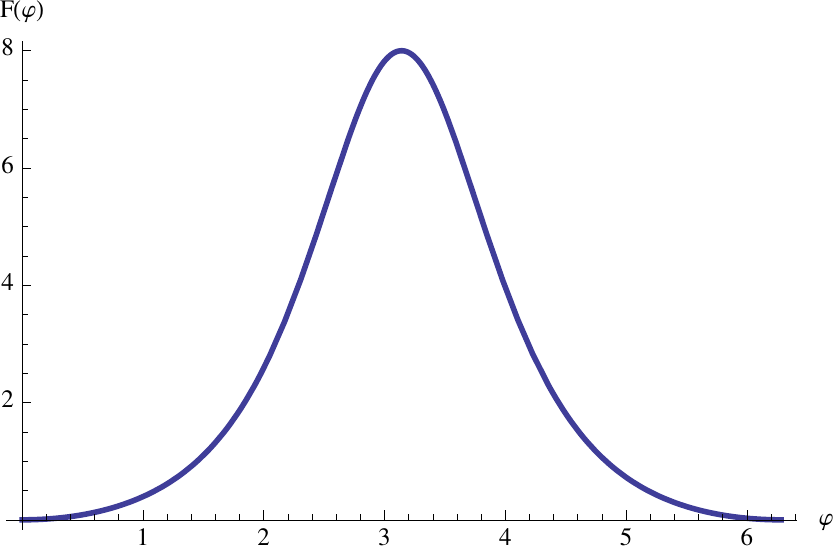}
      \begin{figuretext}\label{Fcminus2.pdf} \upshape
       One period of the function $F(\varphi)$ for $b = |c - 1|$ and $c = -2$.
     \end{figuretext}
     \end{center}
\end{figure}

Assume that $b = |c - 1|$ and $c < -1$. In this case $F(\varphi)$ has double zeros at $\varphi = 2n\pi$, $n \in \Z$, and $F(\varphi) > 0$ away from these zeros, see Figure \ref{Fcminus2.pdf}. Moreover, the set of poles of $F(\varphi)$ is empty. 
An analysis of (\ref{travODE}) shows that corresponding to the interval between two consecutive zeros $2n\pi$ and $2(n+1)\pi$ there exist two smooth solutions $\varphi_1(x)$ and $\varphi_2(x)$ such that
$$\lim_{x \to - \infty} \varphi_1(x) = 2n\pi, \qquad \lim_{x \to \infty} \varphi_1(x) = 2(n+1)\pi.$$
and
$$\lim_{x \to - \infty} \varphi_2(x) = 2(n+1)\pi, \qquad \lim_{x \to \infty} \varphi_2(x) = 2n\pi.$$
This leads to the following result.

\begin{proposition}\label{smoothprop}
For any $c < -1$ and $n \in \Z$, there exists: 
\begin{itemize}
\item A smooth traveling-wave solution $u(x,t) = \varphi(x - ct)$ of (\ref{gsg}) such that 
$$\lim_{x \to - \infty} \varphi(x) = 2n\pi, \qquad \lim_{x \to \infty} \varphi(x) = 2(n+1)\pi.$$

\item A smooth traveling-wave solution $u(x,t) = \varphi(x - ct)$ of (\ref{gsg}) such that
$$\lim_{x \to - \infty} \varphi(x) = 2(n+1)\pi, \qquad \lim_{x \to \infty} \varphi(x) = 2n\pi.$$
\end{itemize}
\end{proposition}
The profiles of these waves for $n = 0$ and $c = -2$ are shown in (a) and (b) of Figure \ref{travfig}.

\subsection{Cusped waves}
In addition to the above smooth traveling waves, equation (\ref{gsg}) also admits cusped traveling waves (cuspons). These arise for $b = |c -1|$ and $-1 \leq c < 1$ (if $c \geq 1$, then $F(\varphi) \leq 0$ for all $\varphi \in \R$, and there do not exist nonconstant solutions of (\ref{travODE})).

\begin{figure}
\begin{center}
    \includegraphics[width=.3\textwidth]{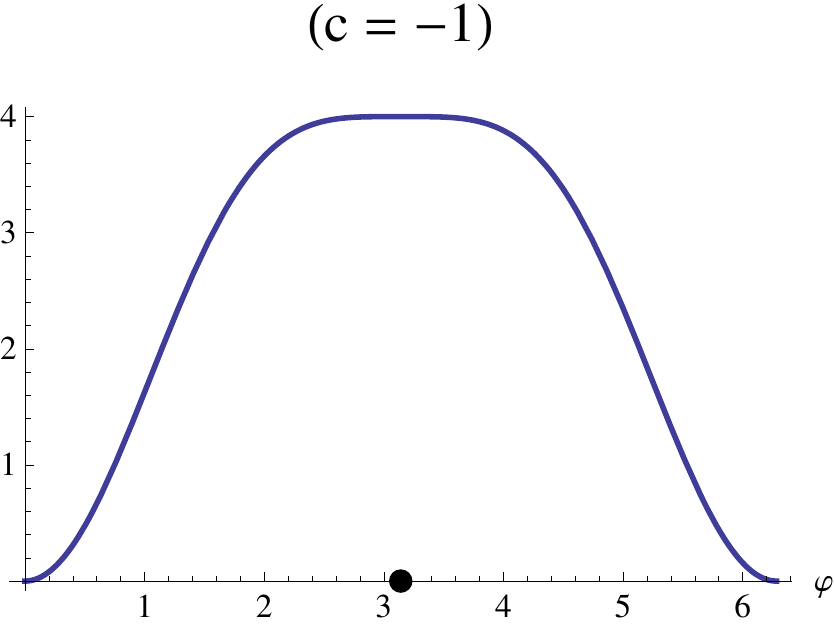} \quad
    \includegraphics[width=.3\textwidth]{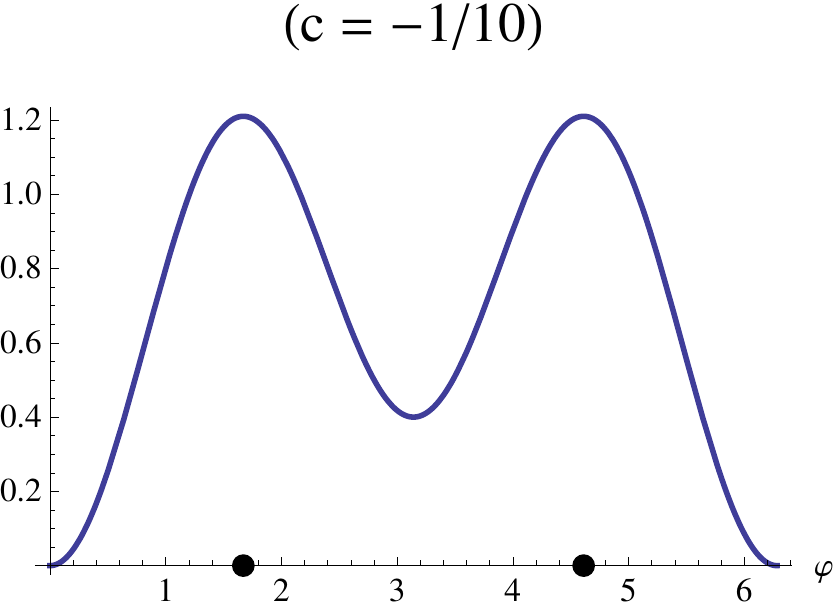} \quad
    \includegraphics[width=.3\textwidth]{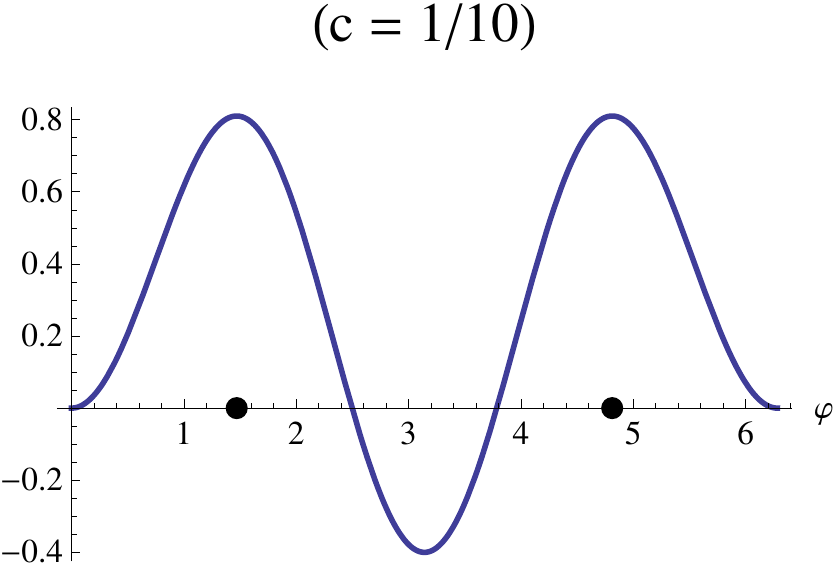} 
     \begin{figuretext}\label{numfig} \upshape
       The numerator $b^2 - (c - \cos(\varphi))^2$ of the function $F(\varphi)$ for $b = |c - 1|$ and $c = -1$ (left),  $c = -1/10$ (middle), $c = 1/10$ (right). The black dots on the $\varphi$-axes denote the values of $\varphi$ at which $F(\varphi)$ has poles.
     \end{figuretext}
     \end{center}
\end{figure}

Assume first that $-1 < c < 0$. In this case $F(\varphi)$ has double zeros at $\varphi = 2n\pi$, $n \in \Z$ and $F(\varphi) > 0$ away from these zeros. See (b) of Figure \ref{numfig} for the graph of the numerator of $F(\varphi)$ when $c = -1/10$. Moreover, $F(\varphi)$ has double poles at $\varphi = \pm \arccos(c) + 2n\pi$, $n \in \Z$. Near one of the poles $\varphi_0 = \pm \arccos(c) + 2n\pi$, the behavior of $F(\varphi)$ is given by
\begin{equation}\label{Fneararccosc} 
 F(\varphi) = \frac{1 - c}{(1 + c) (\varphi - \varphi_0)^2} + O\left(\frac{1}{\varphi - \varphi_0}\right), \qquad \varphi \to \varphi_0.
\end{equation}
Let $\varphi(x)$ be a solution of (\ref{travODE}) such that $\varphi(x_0)$ belongs to the interval $(0, \arccos(c))$ and $\varphi_x(x_0) > 0$ at some point $x_0$. Then, as $x$ increases above $x_0$, the value of $\varphi(x)$ approaches $\varphi_0 = \arccos(c)$. At some point $x_c > x_0$, we have $\varphi(x_c) = \varphi_0$ and near this point integration of (\ref{travODE}) using (\ref{Fneararccosc}) shows that the behavior of $\varphi(x)$ is given by
\begin{equation}\label{nearcusp}
  |\varphi(x) - \varphi_0| =\sqrt{2}\left(\frac{1 -c }{1 + c}\right)^{1/4} \sqrt{|x - x_c|} + O(x - x_c), \qquad x \to x_c.
\end{equation}  
In the case when $\varphi(x) - \varphi_0$ has the same sign locally for $x < x_c$ and $x > x_c$, we say that $\varphi(x)$ has a {\it cusp} at $x = x_c$, i.e. $\varphi(x)$ is continuous, smooth locally on both sides of $x_c$, and 
$$\lim_{x\uparrow x_c} \varphi_x(x) = - \lim_{x \downarrow x_c} \varphi_x(x) = \pm \infty.$$
Cusps such that $\varphi(x)$ has a minimum at $x_c$ are also called {\it anticusps}.
As $x \to \pm \infty$, $\varphi(x)$ decays exponentially to $0$. 
A similar analysis applies to the interval $(-\arccos(c), 0)$ and the translations of these intervals by multiples of $2\pi$. This proves the following result in the case of $-1 < c < 0$: 

\begin{figure}
\begin{center}
    \includegraphics[width=.45\textwidth]{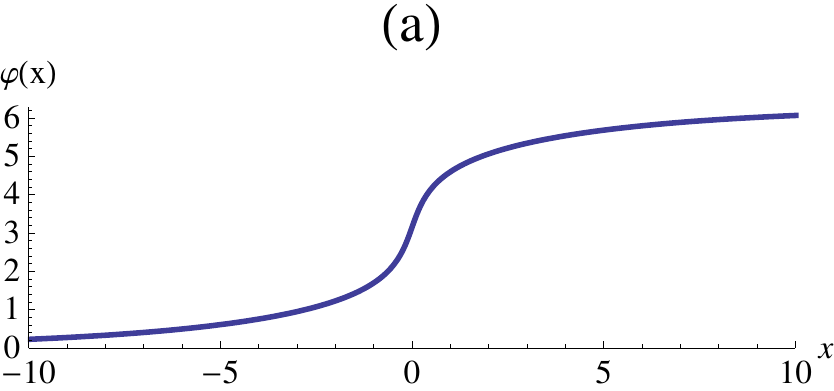} \quad
    \includegraphics[width=.45\textwidth]{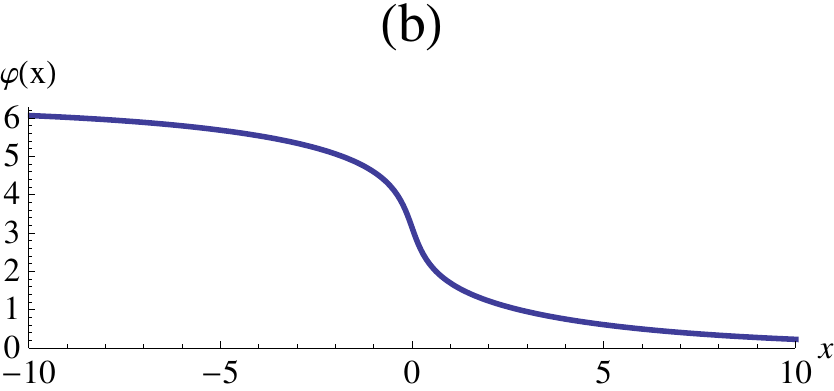} \\ \bigskip
   \includegraphics[width=.45\textwidth]{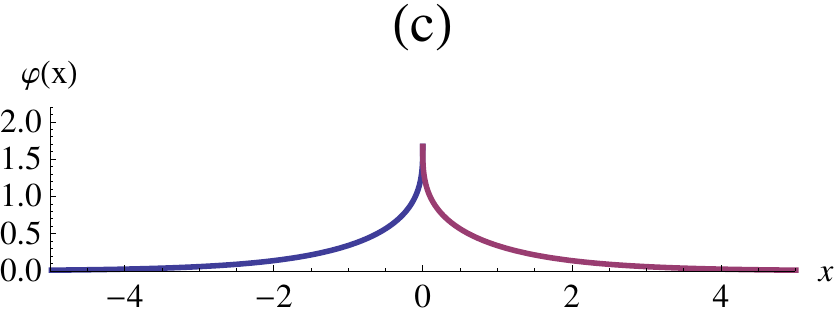} \quad
    \includegraphics[width=.45\textwidth]{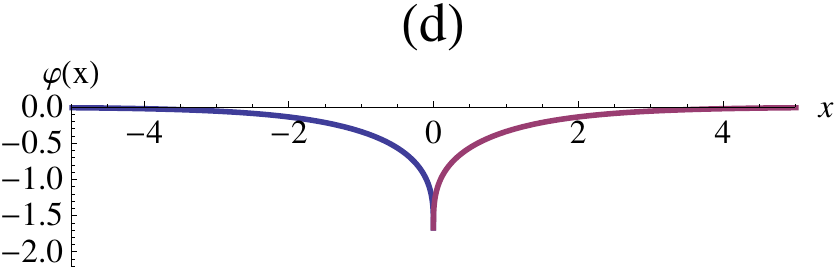} \\\bigskip
   \includegraphics[width=.45\textwidth]{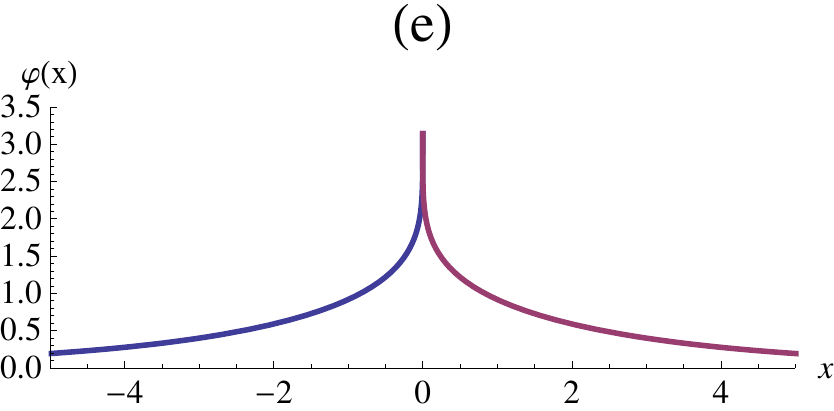} \quad
    \includegraphics[width=.45\textwidth]{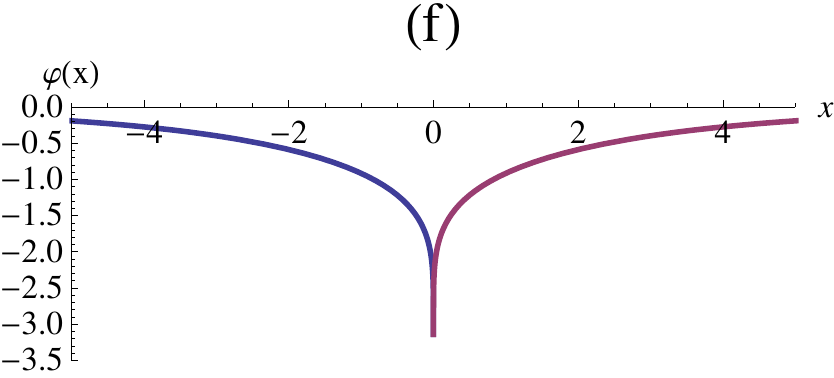} \\\bigskip
   \includegraphics[width=.45\textwidth]{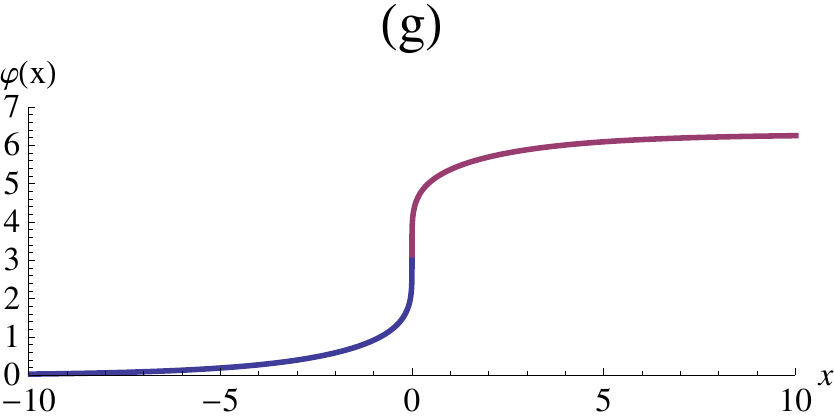} \quad
    \includegraphics[width=.45\textwidth]{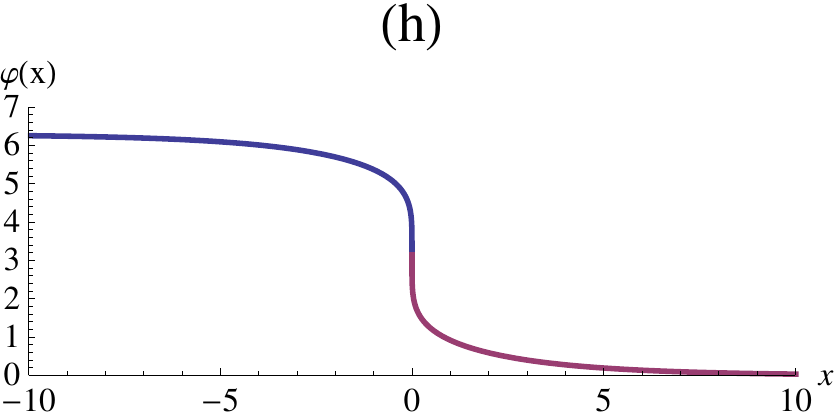} \\
     \begin{figuretext}\label{travfig} \upshape
       Profiles of different types of traveling-wave solutions $u(x,t) = \varphi(x - ct)$ of equation (\ref{gsg}): (a), (b) Smooth traveling waves with $c = -2$; (c) Cusped traveling wave with $c = -1/10$; (d) Anticusped traveling wave with $c = -1/10$; (e) Cusped traveling wave with $c = -1$; (f) Anticusped traveling wave with $c = -1$; (g), (h) Traveling waves with $c = -1$ which combine the cusped and anticusped solutions.
     \end{figuretext}
     \end{center}
\end{figure}

\begin{proposition}\label{cuspprop}
For any $-1 < c < 1$, $c \neq 0$, and $n \in \Z$, there exists: 
\begin{itemize}
\item A cusped traveling-wave solution $u(x,t) = \varphi(x - ct)$ of (\ref{gsg}) such that
$$\lim_{x \to \pm \infty} \varphi(x) = 2n\pi, \qquad \max_{x \in \R} \varphi(x) = 2n\pi + \arccos(c), \qquad \inf_{x \in \R} \varphi(x) = 2n\pi.$$
The maximum is attained at some point $x_c$, and near $x_c$ the behavior of $\varphi(x)$ is given by (\ref{nearcusp}) with $\varphi_0 = 2n\pi + \arccos(c)$, see (c) of Figure \ref{travfig}.

\item An anticusped traveling-wave solution $u(x,t) = \varphi(x - ct)$ of (\ref{gsg}) such that
$$\lim_{x \to \pm \infty} \varphi(x) = 2n\pi, \qquad \min_{x \in \R} \varphi(x) = 2n\pi - \arccos(c), \qquad \sup_{x \in \R} \varphi(x) = 2n\pi.$$
The minimum is attained at some point $x_c$, and near $x_c$ the behavior of $\varphi(x)$ is given by (\ref{nearcusp}) with $\varphi_0 = 2n\pi - \arccos(c)$, see (d) of Figure \ref{travfig}.
\end{itemize}
For $c = -1$ and any $n \in \Z$, there exists: 
\begin{itemize}
\item A cusped traveling-wave solution $u(x,t) = \varphi(x - c t)$ of (\ref{gsg}) such that
$$\lim_{x \to \pm \infty} \varphi(x) = 2n\pi, \qquad \max_{x \in \R} \varphi(x) = (2n +1)\pi, \qquad \inf_{x \in \R} \varphi(x) = 2n\pi.$$
The maximum is attained at some point $x_c$, and near $x_c$ the behavior of $\varphi(x)$ is given by 
\begin{equation}\label{nearcusp2}
  |\varphi(x) - \varphi_0| = 12^{1/3} |x - x_c|^{1/3} + O(|x - x_c|^{2/3}), \qquad x \to x_c.
\end{equation}  
with $\varphi_0 = (2n + 1)\pi$, see (e) of Figure \ref{travfig}.

\item An anticusped traveling-wave solution $u(x,t) = \varphi(x - c t)$ of (\ref{gsg}) such that
$$\lim_{x \to \pm \infty} \varphi(x) = 2n\pi, \qquad \min_{x \in \R} \varphi(x) = (2n - 1)\pi, \qquad \sup_{x \in \R} \varphi(x) = 2n\pi.$$
The minimum is attained at some point $x_c$, and near $x_c$ the behavior of $\varphi(x)$ is given by (\ref{nearcusp2}) with $\varphi_0 = (2n - 1)\pi$, see (f) of Figure \ref{travfig}.
\end{itemize}
\end{proposition}
\proofbegin
It remains to prove the cases $c = -1$ and $0 < c < 1$.
For $0 < c < 1$, $F(\varphi)$ has double zeros at $\varphi = 2n\pi$, $n \in \Z$, simple zeros at $\varphi = \pm \arccos(2c - 1) + 2n\pi$, $n \in \Z$, and double poles at $\varphi = \pm \arccos(c) + 2n\pi$, $n \in \Z$, see Figure \ref{numfig}. 
Since $2c - 1 < c$ for $0 < c < 1$, we have $\arccos(c) < \arccos(2c - 1)$. Thus, $F(\varphi)$ behaves qualitatively in the interval $[0, 2\pi]$ as follows: 
\begin{enumerate}
\item $F(\varphi) > 0$ in the interval $(0, \arccos(c))$. $F(\varphi)$ has a double zero at $\varphi = 0$ and a double pole at $\varphi = \arccos(c)$. 

\item $F(\varphi) > 0$ in the interval $(\arccos(c), \arccos(2c - 1))$. $F(\varphi)$ has a simple zero at $\varphi = \arccos(2c - 1)$. 

\item $F(\varphi) < 0$ in the interval $(\arccos(2c - 1), 2\pi - \arccos(2c -1))$.

\item $F(\varphi) > 0$ in the interval $(2\pi - \arccos(2c - 1), 2\pi - \arccos(c))$. $F(\varphi)$ has a simple zero at $\varphi = 2\pi - \arccos(2c - 1)$ and a double pole at $\varphi = 2\pi - \arccos(c)$.

\item $F(\varphi) > 0$ in the interval $(2\pi - \arccos(c), 2\pi)$. $F(\varphi)$ has a double zero at $\varphi = 2\pi$. 
\end{enumerate}
The proof when $0 < c < 1$ follows from the above properties of the function $F(\varphi)$ and from its $2\pi$ periodicity. Indeed, the solutions listed in the proposition arise in the intervals $(0, \arccos(c))$ and $(2\pi - \arccos(c), 2\pi)$, as well as in the $2\pi$-translations of these intervals. The intervals $(\arccos(c), \arccos(2c - 1))$ and $(2\pi - \arccos(2c - 1), 2\pi - \arccos(c))$ give rise to {\it periodic} cusped traveling waves, which are not included in the proposition.

For $c = -1$, $F(\varphi)$ has double zeros at $\varphi = 2n\pi$, $n \in \Z$, $F(\varphi) > 0$ away from these zeros, and $F(\varphi)$ has fourth-order poles at $(2n + 1)\pi$, $n \in \Z$. The behavior of $F(\varphi)$ near one of the poles $\varphi_0 = (2n + 1)\pi$ is given by
\begin{equation}\label{Fnearpi} 
  F(\varphi) = \frac{16}{(\varphi - \varphi_0)^4} + O\left(\frac{1}{(\varphi - \varphi_0)^3}\right), \qquad \varphi \to \varphi_0.
\end{equation}
It follows that the behavior of a solution $\varphi(x)$ of (\ref{travODE}) near a point $x_c$ at which $\varphi(x_c) = \varphi_0$ is given by (\ref{nearcusp2}). The rest of the proof when $c = -1$ follows as in the previous cases.
\proofend

Since both the cusped and anticusped waves of Proposition \ref{cuspprop} when $c = -1$ satisfy $\varphi \in (2\Z +1)\pi$ at the cusps, it is possible to combine these waves into traveling waves such as those shown in (g) and (h) of Figure \ref{travfig}. These waves are reminiscent of the smooth traveling-wave solutions but have infinite slope at the points where $\varphi \in \pi + 2\pi\Z$.

%In addition to the traveling waves of Proposition \ref{cuspprop}, there also exist time-independent cusped waves $u(x,t) = \varphi(x)$ which arise for $c = 0$. These are given explicitly by
%$$u(x,t) = \pm \arcsin(e^{- |x - x_0|}) + 2n\pi, \qquad n \in \Z, \quad x_0 \in \R.$$

We emphasize that we have only shown formally that the cusped traveling waves are solutions of (\ref{gsg}). In particular, we have not provided an appropriate weak formulation of (\ref{gsg}) of which these waves are solutions. However, in analogy with the CH equation and the form of Eq. (\ref{trav1}), it is expected that such a weak formulation does exist. Indeed, let $\varphi(x)$ be a traveling wave with a cusp located at $x = x_c$. Then $\varphi_{xx}$ will have a singularity at $x_c$. However, since $\varphi_{xx}$ in equation (\ref{trav1}) is multiplied by the factor $\cos(\varphi) - c$ which vanishes at $x_c$, it is possible that the product $\varphi_{xx}(\cos(\varphi) - c)$ is sufficiently regular to allow for a weak formulation. Actually, a direct analog of this mechanism is responsible for the cancellation of the singularities for the peaked and cusped solutions of the Camassa-Holm equation.
If this analogy with CH turns out to be correct, it should also be possible to construct more exotic traveling-wave solutions of equation (\ref{gsg}), such as fractal-like waves and waves with plateaus cf. \cite{Ltrav}.

We have not investigated the stability of the cusped traveling waves of (\ref{gsg}). It is in fact unknown even in the case of the Camassa-Holm and Degasperis-Procesi equations whether the cusped solutions are stable (although the peaked solutions of these equations are known to be stable \cite{C-S, Lpeakstab, LY}).

 \bigskip
\noindent
{\bf Acknowledgement} {\it The authors are grateful to Sergei Sakovich for bringing the references \cite{BRT, Ra, SaSa} to their attention. ASF acknowledges the support of the Guggenheim foundation, USA.}

\bibliography{is}

\end{document}